\documentclass[letter,traditabstract]{aa}
\usepackage{graphicx}
\usepackage{txfonts}
\begin{document}
   \title{A {\em Herschel} PACS and SPIRE study of the dust content of 
    the Cassiopeia~A supernova remnant\thanks{{\em Herschel} is an ESA 
    space observatory with science 
    instruments provided by European-led Principal Investigator consortia 
    and with important participation from NASA.}}

   \author{
 M. J. Barlow\inst{1}
\and
 O. Krause\inst{2}
\and
 B. M. Swinyard\inst{3}
\and
 B. Sibthorpe\inst{4}
\and
 M.-A. Besel\inst{2}
\and
 R. Wesson \inst{1}
\and
 R. J. Ivison\inst{4}
\and
 L. Dunne\inst{5}
\and
 W. K. Gear\inst{6}
\and
 H. L. Gomez\inst{6}
\and
 P. C. Hargrave\inst{6}
\and
 Th. Henning\inst{2}
\and
 S. J. Leeks\inst{3}
\and
 T. L. Lim\inst{3}
\and
 G. Olofsson\inst{7}
\and
 E. T. Polehampton\inst{3,8}
}

   \institute{
Department of Physics and Astronomy, University College London, Gower Street, London WC1E 6BT, UK
\and
Max-Planck-Institut f{\"u}r Astronomie, K{\"o}nigstuhl 17, D-69117 
Heidelberg, Germany 
\and
Space Science and Technology Department, Rutherford Appleton Laboratory, Oxfordshire, OX11 0QX, UK
\and
UK Astronomy Technology Centre, Royal Observatory Edinburgh, Blackford Hill, Edinburgh EH9 3HJ, UK
\and
School of Physics and Astronomy, University of Nottingham, University Park, 
Nottingham NG7 2RD, UK
\and
School of Physics and Astronomy, Cardiff University, The Parade, Cardiff, 
Wales CF24 3AA, UK
\and
Dept of Astronomy, Stockholm University, AlbaNova University Center, 
Roslagstulsbacken 21, 10691 Stockholm, Sweden
\and
Institute for Space Imaging Science, University of Lethbridge, Lethbridge,
Alberta, TJ1 1B1, Canada
}

   \date{}

 
\abstract
{Using the 3.5-m {\em Herschel} Space Observatory, imaging photometry of 
Cas~A has been obtained in six bands between 70 
and 500~$\mu$m with the PACS and SPIRE instruments, with angular 
resolutions ranging from 6 to 37$''$.
In the outer regions of the remnant the 70-$\mu$m PACS image 
resembles the 24-$\mu$m image {\em Spitzer} image, with the emission 
attributed 
to the same warm dust component, located in the reverse shock region. At 
longer wavelengths, the three SPIRE bands are increasingly dominated by 
emission from cold interstellar dust knots and filaments,
particularly across the central, western and southern parts of the 
remnant. Nonthermal emission from the northern part of the remnant 
becomes prominent at 500~$\mu$m. We have estimated and subtracted the 
contributions from the nonthermal, warm dust and cold interstellar dust 
components. We confirm and resolve for the first time a cool 
($\sim$35~K) dust component, emitting at 70-160~$\mu$m, that is located 
interior to the reverse shock region, with an estimated mass of 
0.075~M$_\odot$.}


   \maketitle

\section{Introduction}

The large quantities of dust found in many high-redshift sources (e.g.
Priddey et al. 2003, Bertoldi et al. 2003) have often been interpreted as
having originated in the ejecta of core-collapse supernovae (CCSNe) from
massive stars. Models for CCSNe have predicted the formation of up to
0.1-1~M$\odot$ of dust in their ejecta (e.g. Kozasa et al 1991, Todini \&
Ferrara 2001), which could be sufficient to account for the dust observed
at high redshifts (Morgan \& Edmunds 2003, Dwek et al. 2007) and might
provide a significant source of dust in the local Universe.

Cassiopeia~A (Cas~A), with an age of 330-340 years (Fesen et al. 2006) and 
a distance of 3.4~kpc (Reed et al. 1995), is the youngest known 
core-collapse SNR in the Milky Way, so the mass of swept-up interstellar 
material is much less than that in the ejecta. From optical spectra of 
distant light echoes, Krause et al. (2008) identified it as the product of 
a hydrogen-deficient Type~IIb CCSN. Cas~A has been intensively studied by 
ISO and {\em Spitzer} at infrared wavelengths (e.g. Lagage et al. 
1996; Tuffs et al. 1999; Arendt et al. 1999; Douvion et al. 2001; 
Hines et al. 2004; Ennis et al. 2006; Rho et al. 2008; Smith et al. 
2009). Arendt et al. (1999) derived 0.038~M$_\odot$ of 52~K dust
from a fit to the {\em IRAS} 60- and 100-$\mu$m fluxes, while Rho et al. 
(2008) estimated 0.020-0.054~M$_\odot$ of 65-265~K 
dust to be emitting between 5 and 70~$\mu$m, particularly in a bright ring 
coincident with the reverse shock.
From 450- and 850-$\mu$m SCUBA observations, Dunne et al. (2003) reported 
the presence of excess emission over nonthermal flux levels extrapolated 
from the radio, which they attributed to 2-4~M$_\odot$ of `cold' 
(T$\sim$15-20~K) dust. However, Krause et al. (2004) argued that most of 
the submm excess emission could be due to dust in foreground molecular 
clouds and derived an upper limit of 0.2~M$_\odot$ for cold dust within 
the remnant.
Dunne et al. (2009) reported that the 850-$\mu$m emission from Cas~A was 
polarized at a significantly higher level than its radio synchrotron 
emission and attributed this to $\sim$ 1~M$_{\odot}$ of cold dust or 
alternatively a significantly smaller quantity of iron needles. 
Iron needles were originally proposed by Dwek (2004) and produce a 
very different spectral energy distribution (SED) to `traditional 
grains', with very little flux present at $\lambda < 500~\mu$m. Such 
grains would be consistent with a high polarised fraction.

Nozawa et al. (2010) modelled the evolution of dust in Cas~A and found 
that the observed infrared SED of Cas~A is 
reproduced by 0.08~M$_\odot$ of 
newly formed dust, 0.072~M$_\odot$ of which they inferred to consist of
$\sim$40-K dust in the unshocked regions inside the reverse 
shock. This is supported by recent AKARI and BLAST 65-500-$\mu$m 
photometric observations of Cas~A reported by Sibthorpe et al. (2010). 
Although they concluded that at their longest wavelengths they could not 
isolate any cold dust emission from the SNR from confusing interstellar 
emission, they did however find evidence for a $\sim$33-K `cool' dust 
component, peaking at about 100~$\mu$m, with an estimated mass of 
$\sim$0.06~M$_\odot$.
In the present paper we present new far-IR and submm observations 
of Cas~A obtained with the {\em Herschel} Space Observatory
(Pilbratt et al. 2010).

\section{Observations}

\begin{figure*}
\centering
\includegraphics[width=11.3cm]{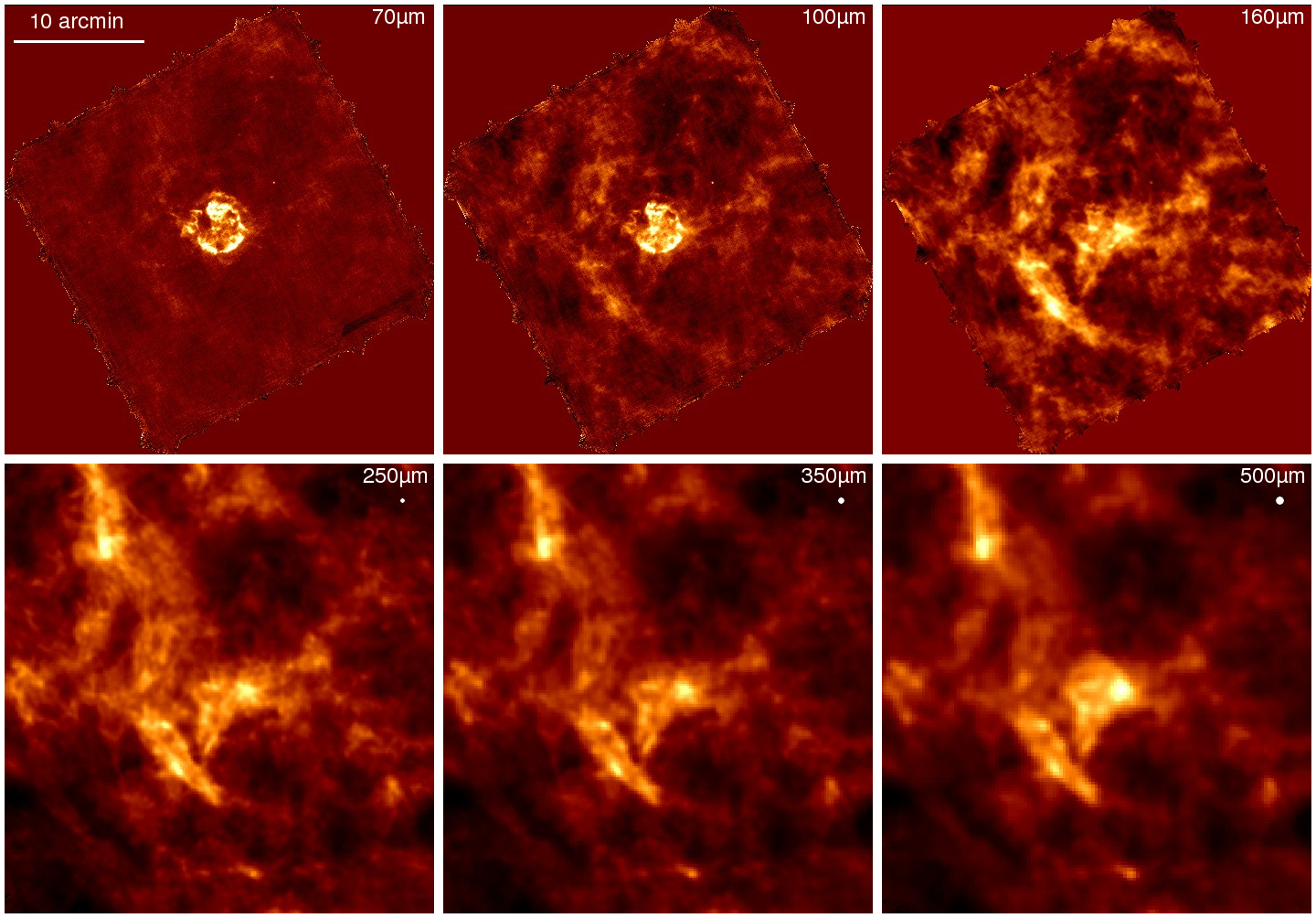}
\caption{Images of Cas~A, obtained in the three PACS bands (top
row) and in the three SPIRE bands (bottom row),
centred at 23h23m26.3s +58$^{\rm o}$48$'$51.33$''$ (J2000.0). North is 
up and east is to the left. The full width half maximum (FWHM) angular
resolutions of the SPIRE images, indicated by the filled white circles
at the top-right of each SPIRE image, are respectively 18, 25 and 
37$''$ at 250, 350 and 500~$\mu$m. With the PACS scan-map speed of
20$''$/sec the FWHM resolutions were 5.8, 7.8 and 12.0$''$ at 70, 100
and 160~$\mu$m, respectively.
}
\end{figure*}

Cas~A was observed with the SPIRE imaging photometer on 2009 Sep 12 and 
Dec 17. The SPIRE instrument and its in-orbit performance are 
described by Griffin et al. (2010), and the SPIRE astronomical calibration 
methods and accuracy are outlined by Swinyard et al. (2010). The 
photometer's absolute flux calibration uncertainty is estimated to be 
15\%. On each occasion scan maps covering a $32'\times32'$ area 
centred on Cas~A were obtained simultaneously at 250, 350 and 500~$\mu$m, 
with an on-source integration time of 2876~s. 
The remnant was observed with the PACS imaging photometer on 2009 Dec 17. 
The PACS instrument, its in-orbit performance and calibration are 
described by 
Poglitsch et al. (2010); the absolute flux calibration uncertainty of the 
photometer is estimated to be 20\%. Scan maps comprised of two orthogonal
scan legs, each of 22$'$ in length, were obtained using the
70+160-$\mu$m and 100+160-$\mu$m channels. For each pair of filters
the on-source integration time was 2376~s.

A montage showing the images in the six PACS and SPIRE bands is presented 
in Fig.~1. The PACS 70-$\mu$m image (Fig. 1, top-left and Fig. 2, 
top-right) strongly resembles the similar angular resolution {\em Spitzer} 
24-$\mu$m MIPS image (Hines et al. 2004); a bright ring of warm dust 
emission is coincident with the reverse shock, while the fainter outer 
emission edge coincides with the forward shock. In the longer wavelength 
images, knots and lanes of diffuse interstellar dust emission envelope the 
SNR - this emission is particularly bright at the central, western and 
southern parts of the remnant, where its morphology closely matches that 
of molecular line maps, such as the $^{13}$CO emission map presented by 
Wilson \& Batrla (2005). In the SPIRE 500-$\mu$m image, the nonthermal 
emission from the northern parts of the remnant becomes prominent,
coincident with emission seen with SCUBA at 850~$\mu$m (Dunne et al. 
2003).

The second row of Table~1 lists the total flux density measured from 
Cas~A in each of the six {\em 
Herschel} bands, using a 165$''$ radius aperture that should 
encompass everything within the forward shock region,
located at 153$\pm$12$''$ (Gotthelf et al. 2001). These total
flux densities were measured relative to four `floor' regions
located to the north and southwest of the nebula. The total
flux densities listed for Cas~A include the emission from 
the cold interstellar 
dust that is superposed on the remnant. The first row of Table~1 lists
previously published flux densities for Cas~A at wavelengths in common.
The SPIRE 500-$\mu$m flux overlaps the SCUBA and BLAST 
450/500-$\mu$m fluxes (Dunne et al. 2003; Sibthorpe et al. 2010)
but at shorter wavelengths the PACS and SPIRE flux densities are factors 
of 1.6-2.2 times larger than published values from {\em IRAS} and {\em 
Spitzer} (Hines et al. 2004) and from {\em AKARI} and BLAST 
(Sibthorpe et al. 2010) that are listed in the first row of Table 1. We 
attribute these differences to the fact that the higher angular 
resolution of {\em Herschel} enabled lower `floor' points in 
\begin{figure*}
\centering
\includegraphics[width=11.3cm]{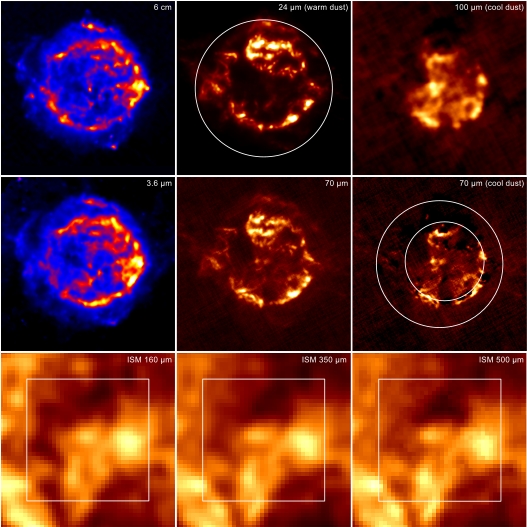}
\caption{Images of Cas~A at infrared, submillimetre and radio 
wavelengths.
The top six images are 7$'$ on a side, while the lower three
images are 10$'$ on a side, with inset boxes showing the 7$'$ 
field. North is up and east is to the left. 
The inner and outer circles in the middle-right image respectively show 
the positions of the reverse and forward shocks according to Gotthelf et 
al. (2001), while the 165\arcsec-radius circle in the top-middle 
image encloses the area for which the fluxes listed in Table~1 were
measured. See text for further details.}
\end{figure*}
the diffuse background emission to be resolved and subtracted, whereas the 
lower angular resolutions of smaller aperture telescopes, at 
wavelengths in common, is likely to have caused higher mean background 
levels to be estimated. 

\begin{table}
\caption{Total and individual component flux densities (in Jy) for Cas~A}
\label{fluxes}
{\tiny
\begin{tabular}{lcccccc}
\hline
 & 70$\mu$m & 100$\mu$m & 160$\mu$m & 250$\mu$m & 350$\mu$m & 500$\mu$m \\
\hline
Published & 107$^1$ & 105$^2$ & 101$^2$ & 76$^3$& 49$^3$ & 42$^3$,70$^4$ \\
      & $\pm$22 & $\pm$21 & $\pm$20 & $\pm$16 & $\pm$10 & $\pm$8,$\pm$16 \\
{\em Herschel} & 169 & 192 & 166 & 168 & 92 & 52 \\
         & $\pm$17 & $\pm$19 & $\pm$17 & $\pm$17 & $\pm$10 & $\pm$7 \\
Nonthermal & 6.3 & 8.1 & 11.2 & 15.4 & 19.4 & 24.9 \\
           & $\pm$0.6 & $\pm$0.7 & $\pm$0.9 & $\pm$1.1 & $\pm$1.3 & $\pm$1.6 \\
Warm dust & 120 & 63 & 22 & 7.0 & 3.1 & 1.2 \\
 & $\pm$12 & $\pm$6 & $\pm$2 & $\pm$0.8 & $\pm$0.4 & $\pm$0.2 \\
Cold IS dust & 18 & 76 & 123 & 141 & 69 & 27.5 \\
 & $\pm$4 & $\pm$11 & $\pm$17 & $\pm$17 & $\pm$10 & $\pm$7 \\
Cool dust & 25 & 29$^5$ & 10 & 4.6 & 0.5 & -1.6 \\
 & $\pm$7 & $\pm$11 & $\pm$17 & $\pm$17 & $\pm$10 & $\pm$4 \\
\hline
\end{tabular}
}
{\tiny $^1$MIPS (Hines et al. 2004); $^2$AKARI 90 and 170~$\mu$m 
(Sibthorpe et 
al. 2010); $^3$BLAST 250, 350 and 500~$\mu$m (Sibthorpe et al. 2010); 
$^4$ SCUBA 450~$\mu$m (Dunne et al. 2003);
$^5$After subtracting a line contribution of 16~Jy (see Sect.~3).}
\end{table}

\section{Emission component decomposition}

In order to investigate the `cool' dust emission component in Cas~A 
that was diagnosed by Sibthorpe et al. (2010) from an analysis 
of their AKARI and BLAST data, we have followed a similar 
procedure by attempting to identify and subtract the contributions made at 
each wavelength by (a) the remnant's nonthermal (synchrotron) emission; 
(b) the warm dust component that dominates the {\em Spitzer} 24-$\mu$m 
image; and (c) the cold interstellar dust component. In addition, we have 
estimated the contributions made by line emission to the PACS in-band 
fluxes.
\newline
{\bf The nonthermal component:} We extracted from the archive and 
reprocessed a 6-cm VLA dataset on Cas~A obtained in 1997/8,
convolving it to 6$''$ resolution, as shown in Fig.~2 (upper 
left). Also shown in Fig.~2 is the 3.6-$\mu$m IRAC image 
obtained by Ennis et al. (2006), which was also convolved to a 
resolution of 6$''$; as Ennis et al. noted, the morphologies of the 
6-cm and 3.6-$\mu$m images correspond very closely, indicating that both 
are dominated by the nonthermal emission component. We corrected the 
3.6-$\mu$m image for extinction based on the X-ray absorption results of 
Willingale et al. (2002) and ratioed the two nonthermal images to produce 
a spectral index map which was quite smooth, yielding a mean spectral 
index of $-0.70\pm0.05$. We therefore adopted a spectral index of $-0.70$ 
to estimate the remnant's nonthermal emission in each of the six PACS and 
SPIRE bands (third row of Table~1) and to generate images convolved to 
the resolution of each of the bands. The other images in Fig.~2
have had the appropriate nonthermal component image subtracted.
\newline
{\bf The warm dust component:} Fig.~2 (middle-top panel) shows the 
24-$\mu$m MIPS image obtained by Hines et al. (2004). They noted the 
similarity between the 24- and 70-$\mu$m MIPS images, pointing to a 
common emitting component, which we term the `warm dust' component, 
peaking in a bright ring coincident with the position of the reverse 
shock. Our PACS 70-$\mu$m image (Fig.~2, central panel) has a similar 
resolution to the MIPS 24-$\mu$m image and shows a similar outer 
morphology, but more emission is evident from the remnant's interior in 
the 70-$\mu$m image. We therefore normalised the MIPS 24-$\mu$m image to 
the surface brightness levels in the outer parts of the remnant in the 
PACS 70-$\mu$m image and subtracted it, to obtain the difference image 
shown in the middle-right panel of Fig.~2. The total `warm dust' 
contribution at 70~$\mu$m, obtained from the scaled-up 24-$\mu$m image, is 
120$\pm$6~Jy. We extrapolated this warm dust component from 70~$\mu$m to 
longer 
wavelengths using the predicted spectrum from $3\times10^{-3}$~M$_\odot$ 
of 82-K magnesium protosilicate, found by Hines et al. (2004) to fit the 
24-70-$\mu$m MIPS spectrum, in order to obtain the flux densities listed 
in the 4th row of Table~1. Warm dust component images, convolved to the 
appropriate angular resolutions, were subtracted from the images obtained 
at 100~$\mu$m and longwards, as were the appropriate nonthermal images, 
before estimating and subtracting the contribution from superposed 
cold interstellar dust, discussed next.
\newline
{\bf The cold interstellar and cool Cas~A dust components:} The bottom row
of Fig.~2 shows the 160, 350 and 500-$\mu$m images of Cas~A
after subtracting scaled images of the other components.
For illustration purposes they are shown convolved
to the same 37$''$ resolution as the 500-$\mu$m image. 
These residual images show a strikingly similar morphology,
indicating that they are emitted by the same cold interstellar
dust particles. To obtain these maps in an iterative way we started 
with maps corrected for the nonthermal and warm components and
determined average 100/160 and 70/160-$\mu$m
flux ratios for several bright regions located outside
the remnant. We then applied these ratios to the 160-$\mu$m 
image and subtracted them from 70- and 100-$\mu$m images which 
had been convolved to the 160-$\mu$m resolution. A consistent 
cool dust morphology is seen in the resulting 70 and 100-$\mu$m 
`cool dust' images (Fig.~2, middle-right and top-right). Note that in 
this first step, the 160-$\mu$m map initially
still contained a contribution from the cool SN dust component.
In order to determine its contribution at 160~$\mu$m, we subtracted
a scaled image of the cool component at 70~$\mu$m (where the
ISM and line contamination is smallest) from the
160-$\mu$m map iteratively, until its visible imprint 
was minimized. This corrected 160-$\mu$m ISM map was then used
iteratively to obtain more accurate 70- and 100-$\mu$m images
of the cool dust component.
Our estimates for the flux densities in each band from the cold 
interstellar dust emission that is superposed on the remnant are listed 
in the penultimate row of Table~1. We note that the relative uncertainties
of individual emission components are smaller than the absolute 
calibration uncertainties associated with the total flux densities.
\newline
{\bf Emission line contributions to the PACS bands:} Archival ISO-LWS 
43-197-$\mu$m grating spectra, obtained with an aperture size of 80$''$, 
exist for six positions across Cas~A, and for one offset 
position (see Fig.~4 of Docenko \& Sunyaev 2010). The spectra show strong 
broad emission from the [O~{\sc i}] 63-$\mu$m and [O~{\sc iii}] 52- and 
88-$\mu$m lines. After convolving with the filter and instrumental 
response functions, the line contribution to the 70- and 160-$\mu$m
bands was found to be negligible but the 88-$\mu$m line was 
found to make a $\sim$16~Jy contribution to the PACS 
100-$\mu$m band - this has been subtracted to give the 100-$\mu$m `cool 
dust' flux density listed in the last row of Table~1. 
The spectral energy distributions of each of the emitting components are
plotted in Fig.~3 in the online Appendix.

\section{Discussion: the mass of cool dust in Cas~A}

Following subtraction of the nonthermal, warm dust and cold interstellar
dust components, the 100-$\mu$m image shown in Fig.~2 (top-right) shows a
similar morphology to the cool dust 70-$\mu$m image shown below it. These
represent the first resolved images of this dust component, whose
existence was also inferred by Tuffs et al. (2005; 60-200-$\mu$m ISOPHOT)
and Sibthorpe et al. (2010; 65-500-$\mu$m AKARI/BLAST). The flux densities
in each band from the cool dust component are listed in the final row of
Table~1. They can be fitted (Fig.~3) by 0.075$\pm$0.028~M$_\odot$ of
35$\pm$3-K $\lambda^{-2}$ emissivity silicate dust having a 160-$\mu$m
absorption coefficient of 9.8~cm$^2$~g$^{-1}$ (Dorschner et al. 1995).
Sibthorpe et al. derived a 33-K cool dust mass of 0.055~M$_\odot$
(0.066~M$_\odot$ with the dust absorption coefficients used here),
consistent with our own estimate.

Nozawa et al. (2010) modelled the Hines et al. (2004) 8-100-$\mu$m SED of
Cas~A with 0.008~M$_\odot$ of shock-heated warm dust and 0.072~M$_\odot$
of unshocked cool dust in the remnant's interior. Their dust formation
model for the Cas~A ejecta predicted 0.17~M$_\odot$ of new dust, from
which they suggested that 0.09~M$_\odot$ had already been destroyed by the
reverse shock. If the 0.075~M$_\odot$ of cool interior dust that we find
here is to survive its 5000~km~s$^{-1}$ encounter with the reverse shock,
it will need to be protected by being inside very dense clumps. If most of
the dust was eventually destroyed, then remnants of this type 
would not make a significant contribution to the dust content of the ISM, 
and could even dilute it.

The present observations provide no direct evidence for the presence
of significant quantities of cold ($<$25~K) dust within Cas~A - the
500-$\mu$m emission that is visible from the least obscured, northern, 
part of the remnant is, to first order, removed when the predicted
nonthermal emission contribution is subtracted (Fig.~2; bottom-right). 
The cause of the 850-$\mu$m excess in the SCUBA map
of the northern part of the remnant is therefore unresolved. 
Spectroscopic observations are planned with PACS and SPIRE for various 
on- and off-remnant positions. These should enable a clean separation of
line contributions and a full sampling of the continuum energy
distribution from 50-670~$\mu$m.

\begin{acknowledgements}

{\bf PACS} has been developed by a consortium of institutes led by MPE
(Germany) and including UVIE (Austria); KUL, CSL, IMEC (Belgium); CEA,
OAMP (France); MPIA (Germany); IFSI, OAP/AOT, OAA/CAISMI, LENS, SISSA
(Italy); IAC (Spain). This development has been supported by the funding
agencies BMVIT (Austria), ESA-PRODEX (Belgium), CEA/CNES (France),
DLR (Germany), ASI (Italy), and CICT/MCT (Spain).
{\bf SPIRE} has been developed by a consortium of institutes led by
Cardiff Univ. (UK) and including Univ. Lethbridge (Canada);
NAOC (China); CEA, LAM (France); IFSI, Univ. Padua (Italy);
IAC (Spain); Stockholm Observatory (Sweden); Imperial College London,
RAL, UCL-MSSL, UKATC, Univ. Sussex (UK); Caltech, JPL, NHSC,
Univ. Colorado (USA). This development has been supported by
national funding agencies: CSA (Canada); NAOC (China); CEA,
CNES, CNRS (France); ASI (Italy); MCINN (Spain); SNSB
(Sweden); STFC (UK); and NASA (USA). 
These observations were taken as part of the Science Demonstration Phase of
the Mass-loss of Evolved StarS (MESS) Guaranteed Time Key Programme.
(Groenewegen et al., in prep.)

\end{acknowledgements}

\vspace{0.9cm}
\noindent
{\bf \Large Online Appendix}

\begin{figure*}
\centering
\includegraphics[width=15.3cm]{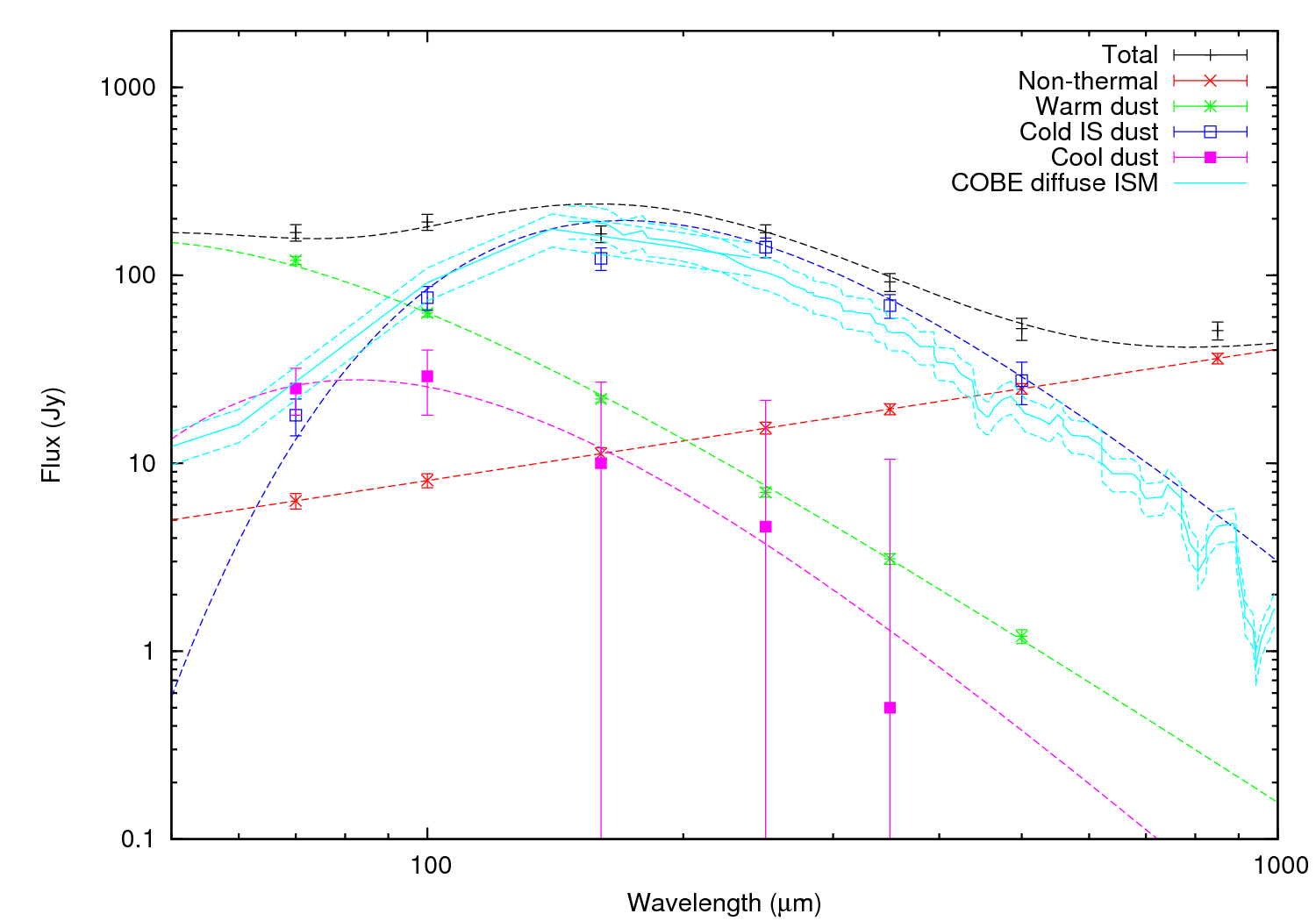}
\caption{The derived 70-850-$\mu$m spectral energy distributions of the 
components contributing to the observed emission from Cas~A.
See Sect.~3 for a description of how the flux densities from each
component at each wavelength were estimated.
Red: nonthermal flux densities estimated from a power-law fit between the 
6-cm and 3.6-$\mu$m flux densities; Green: warm dust component flux 
densities; 
Blue: flux densities for the cold interstellar dust component. Also
shown are a $\lambda^{-2}$ emissivity 17-K fit to the cold
IS dust flux densities (blue dashed line), and a comparison with
the Dwek et al. (1997) COBE DIRBE/FIRAS average ISM spectral
energy distribution (turquoise; shown with the quoted
$\pm$20\% 1$\sigma$ uncertainty limits). Magenta: the derived
flux densities for the cool dust component, together with a 35-K 
$\lambda^{-2}$ emissivity fit (dashed magenta line). The blue points 
show the PACS and SPIRE 70-500-$\mu$m total flux densities measured for 
Cas~A, as well as the SCUBA 850-$\mu$m flux density measured by Dunne et 
al. (2003). The black dashed line corresponds to the sum of the fits to
the nonthermal, warm dust, cool dust and cold IS dust components.}
\end{figure*}

\end{document}